\documentclass[preprint,12pt]{elsarticle}

\usepackage{caption}
\usepackage{subcaption}

\usepackage[cmex10]{amsmath,mathtools}
\usepackage{mdwmath}
\usepackage{mdwtab}
\usepackage{amssymb}

\usepackage{algorithmic}
\usepackage{algorithm}
\usepackage{picins}
\usepackage{float}

\usepackage{epstopdf}
\usepackage{multirow}
\usepackage{color}

\journal{XXXXXX}

\begin{document}

\begin{frontmatter}

\title{SenseFlow: An Experimental Study for Tracking People}

\author[sutd]{Kai Li}
\ead{kai\_li@sutd.edu.sg}

\author[sutd]{Chau Yuen}

\author[unsw]{Salil S. Kanhere}

\author[peking]{Kun Hu}

\author[peking]{Wei Zhang}

\author[peking]{Fan Jiang}

\author[peking]{Xiang Liu}

\address[sutd]{SUTD-MIT International Design Center, The Singapore University of Technology and Design, Singapore}

\address[unsw]{School of Computer Science and Engineering, The University of New South Wales, Sydney, Australia}

\address[peking]{School of Software and Microelectronics, Peking University, Beijing, China}
            
\begin{abstract}
The main challenges in large-scale people tracking are the recognition of people density in a specific area and tracking the people flow path. To address these challenges, we present \textit{SenseFlow}, a lightweight people tracking system. \textit{SenseFlow} utilises off-the-shelf devices which sniff probe requests periodically polled by user's smartphones in a passive manner. 
We demonstrate the feasibility of \textit{SenseFlow} by building a proof-of-concept prototype and undertaking extensive evaluations in real-world settings. 
We deploy the system in one laboratory to study office hours of researchers, a crowded public area in city to evaluate the scalability and performance ``in the wild", and four classrooms in the university to monitor the number of students. 
We also evaluate \textit{SenseFlow} with varying walking speeds and different models of smartphones to investigate the people flow tracking performance. 
\end{abstract}


\end{frontmatter}

\section{Introduction}
\label{intro}
In recent years, the popularity of mobile and pervasive computing stimulates many research efforts on wireless  people tracking. An increasingly common requirement of people tracking is to extract information regarding the people density and moving trajectories in an environment~\cite{li2015senseflow,teixeira2010survey}. Many questions could be asked, e.g.: how many customers visit a shopping mall everyday and which shops get more customers than the others in the mall? How many people are waiting in a subway station, and how the flows of people move inside interchange stations? How long do the people walk from an entrance to an exit? Such tracking information can help the service provider understand public space usage patterns so as to improve their resource allocation~\cite{petrenko2013human}. Moreover, it is also a significance to understand pedestrian flows~\cite{kjaergaard2012mobile} and human mobility~\cite{xu2012exploiting}, e.g., for social psychology studies to sense people's mood based on their attitude towards joining flocks or for epidemiological studies to consider how often people join flocks~\cite{rachuri2010emotionsense}. Furthermore, flock detection can also enable new tools for emergency research studies which is concerned with the size of flocks and how they form, dissolve or are slowed down by constraints, e.g., door passages~\cite{schadschneider2011evacuation}. 

The integration of wireless sensing techniques and mobile devices such as smartphones is enabling next generation light-weight people tracking applications~\cite{lane2010survey}. A possible way of people tracking is to utilise probe requests that are broadcast by smartphones for Wi-Fi connection~\cite{fujinami2013human,musa2012tracking} as a proxy for the people present in the area. People's trajectory can be tracked only when these spatial-temporal probe requests at different locations are fully collected by the sensors. Unfortunately, translating this broad idea into a practical people tracking system entails a range of challenges. 
First, a large number of probe requests (i.e. hundreds of people and smartphones in a crowded area) is generated in real time. Forwarding all the probe requests generates a peak of traffic in the network, which poses a challenging problem on the data collection and processing. 
Second, the probe request generating interval highly depends on operational mode of a mobile device (shown by our experiments in Section~\ref{expBeacon}). As a result, the smartphone will not be tracked if its probe request is missed when the user moves across the sensing range of the sensor. 
Third, tracking people flow in a spacious area is non-trivial since multiple trajectories are available between any two locations, and the exact people movement pattern is often unknown. Additionally, the probability of probe request detection decreases with pedestrian walking speed since the sensor node has a limited sensing coverage. However, no existing work studies how the probe request frequency and human walking behaviour effect people flow tracking performance. 

\begin{figure}[htb]
\centering
\includegraphics[width=0.8\textwidth]{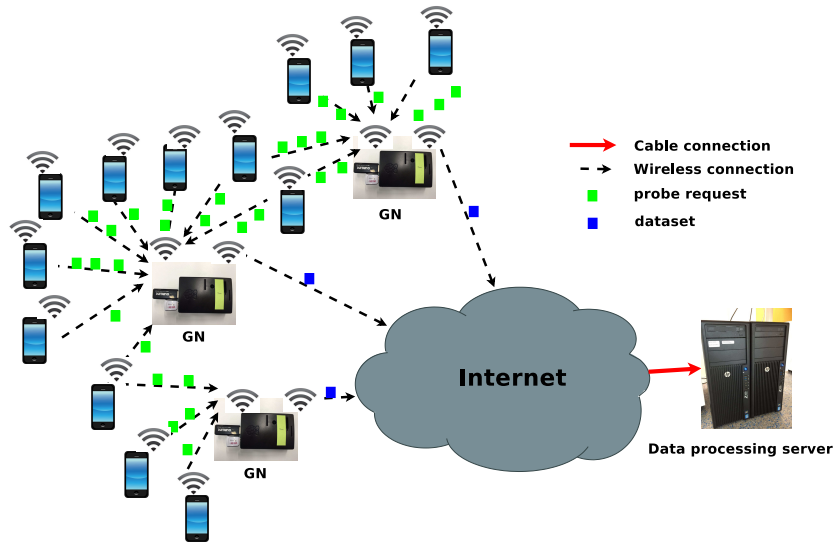}
\caption{System architecture of \textit{SenseFlow}. Each GN covers a specific area and the datasets of GN are uploaded to the sever through public network.}
\label{fig_arc}
\end{figure}

To address the above issues, in this paper, we propose \textit{SenseFlow}, a lightweight sensing testbed to monitor people quantity in a given area and track people flow movement, based on a passive collection of the probe requests from their smartphones without knowing the environmental feature or fingerprint. Specifically, \textit{SenseFlow} uses a number of time-synchronised wireless gateway nodes (GNs) to collect probe requests broadcasting by users' smartphones at different locations, as shown in Figure~\ref{fig_arc}. 
A user's presence is detected when the probe request of the smartphone is received by the GN the user passes by. Therefore, the trajectory of the user is obtained by tracking the probe requests on a series of GNs according to time. In \textit{SenseFlow}, a server is deployed to amalgamate the data from all GNs so as to investigate when and which GNs collect the probe requests. 
To provide a real-time people tracking, one of data collection schemes requires the GN to transmit all probe requests from each smartphone. However, transmitting all probe requests in real time generates a large network traffic. To address this issue, we propose a novel Probe Request Interval-based Data Collection Scheme (PRI-DCS), where the GN extracts source MAC address and timestamp from probe request, and upload the extracted data (dataset) according to the probe request interval of smartphone. 
The datasets transmitted from all GNs are amalgamated to monitor people density of the area by counting the amount of MAC addresses during a certain time interval. 

Furthermore, we find that the people tracking system based on smartphone monitors people density inaccurately due to a probe request overhearing problem, where the nodes deployed at adjacent locations can receive probe requests from the same smartphone. To address this issue, in \textit{SenseFlow}, we extend the PRI-DCS by selecting the GN that has the maximum Received Signal Strength Indicator (RSSI) of probe requests as the location where the smartphone presents. 

We formulate the trajectory of individual user as a sequence of GN IDs in the datasets. Next, we implement a Longest Common Subsequence (LCS) algorithm to recognise the user's trajectory, and track the flow of people from one a specific starting point to an ending point. 

The rest of this paper is organised as follows. Section \ref{relatedwork} presents the related work on different types of people tracking system. 
We then present the details of the proposed \textit{SenseFlow} system in Section \ref{system}. 
Section \ref{experiment} illustrates the implementation and evaluation results in both controlled and real-world experiments. 
Finally, we conclude this work in Section \ref{cond}.

\section{Related Work}
\label{relatedwork}
In this section, we present a brief overview of previous work of people tracking systems covering the people density monitoring and movement tracking. 

\subsection{Sensor-based People Tracking Systems}
The GPS-based localisation system is widely used for outdoor position determination and this technology is currently implemented in many mobile devices~\cite{deak2012survey}. Unfortunately, the main challenge in indoor environments is the unavailability of GPS signals since the technology requests for Line-of-Sight when connecting to satellites. In addition, such system requires the user to install an application on the mobile device in order to enable GPS localisation, which does not track people in a passive way. 

Camera-based system was proposed to address the people tracking using thermal infrared, stereo and time of flight camera~\cite{chen2013survey,fernandez2010optical}. The vast majority of human-detection approaches currently deployed in camera-based systems rely on background subtraction, pattern matching and face recognition, which can process the conventional images from the camera. However, these systems are affected by lighting variations and shadows. Moreover, camera-based system has limited coverage due to a fixed location and angle~\cite{guvensan2011coverage}. 

Apart from cameras, other devices used for people tracking are range finders, such as radar, and sonar. In~\cite{mozos2010multi}, Mozos et al. proposes people tracking by using multiple layers of 2D laser range scans. \cite{premebida2009exploiting} presents a valuable analysis of pedestrian detection in urban scenario using exclusively lidar-based features. Unfortunately, the impressionable wave and laser signal lead a large number of false negatives~\cite{teixeira2010survey}. 

Sensor fusion approaches build upon the use of multiple sensors such as camera and microphone~\cite{beal2003graphical}, camera and range finder~\cite{hu2012bayesian}, camera and motion sensor~\cite{zhaivm}, etc. The idea is to combine their advantages while cancelling out their disadvantages as much as possible. Unfortunately, sensor fusion systems require the installation of a complex infrastructure, which causes a large capital investment in setup~\cite{teixeira2010survey}. In addition, the state-of-the-art sensor fusion systems can hardly meet the accuracy and delay requirement for large-scale people tracking. 

The comparison of people tracking systems in literature is summarised in Table~\ref{tb_methods}. 

\begin{table}[htb]
    \centering
    \caption{Comparison of people tracking systems.}
    \begin{tabular} {|p{2.3cm}|p{1.5cm}|p{1.5cm}|p{1.5cm}|p{1.5cm}|p{2.0cm}|} \hline
        \bf{} & \bf{Camera} & \bf{GPS} & \bf{Range finders} & \bf{Sensor fusion} & \bf{SenseFlow} \\ \hline
	Indoor tracking  & Yes & No & Yes & Yes & Yes\\ \hline
	Large-scale people tracking & No & Yes & No & Yes & Yes\\ \hline	
        	Trajectories recognition accuracy  & Low & High & Low & Low & High\\ \hline
        	Tracking latency  & High & Low & Low & High & Low\\ \hline 
        	System Complexity  & High & Low & Low & High & Low\\ \hline 	
   \end{tabular}
\label{tb_methods}
\end{table}

\subsection{People Tracking with Smartphone}
Wi-Fi RSSI~\cite{sigg2014passive} and ambience fingerprinting~\cite{azizyan2009surroundsense} have been researched for mobile device based indoor localisation. In~\cite{sigg2014passive}, Wi-Fi RSSI in received packets at a mobile phone is utilised to detect user's presence in the area. Furthermore, fluctuation on WiFi RSSI might indicate the number of people around or even activities conducted in proximity. 
By combining optical, acoustic, and motion attributes from various sensors of the mobile phone, \textit{SurroundSense} system constructs an identifiable fingerprint, which includes ambient sound, light, colour, RF and user movement~\cite{azizyan2009surroundsense}. This fingerprint is then used to identify the user's location. 

In~\cite{barbera2013signals}, the social links in a venue of large political and religious gatherings are studied from the probe requests. A database that associates each device is built, as identified by its MAC address, to the list of SSIDs derived from its probe requests. Moreover, an automated methodology is proposed to learn the social links of mobile devices given that two users sharing one or more SSIDs indicate a potential social relationship between the two. 

A tracking system, which consists of a number of road-side Wi-Fi monitors and a central server, is presented in~\cite{musa2012tracking}. They propose a probabilistic method to estimate smartphone trajectories for single user from Wi-Fi detections. It is shown that the accuracy of Wi-Fi tracking depends to a large degree on the density and geometry of monitors' deployment. 

A low-cost tracking system for pedestrian flow estimations based on Bluetooth and Wi-Fi captures is proposed in~\cite{schauer2014estimating}. The system tracks a pedestrian's movement through an area of interest by capturing the device specific MAC address at different monitor nodes located at the entrances/exits to this area of interest. Furthermore, they propose a hybrid tracking approach that considers both the RSSI value and the time when a MAC address was captured. 

Compared with the existing solutions and concepts, we experimentally measure the probe request interval with different operational modes of three typical smartphone operating systems, and the effect of human walking behaviour. With those studies, our people tracking system collects dataset packets from GNs based on the probe request interval to mitigate network traffic and tracking latency. Moreover, our approach is able to address the probe request overhearing problem to improve the tracking accuracy. 

\section{SenseFlow System}
\label{system}
In this section, we first present the design of gateway node and system architecture with a new data collection scheme. Next, we study the probe request overhearing problem in the people tracking system. We then outline a people tracking algorithm based on the datasets from GNs.

\subsection{Probe Request Collection and Overhearing}
\label{architecture}
\begin{figure}[htb]
\centering
\includegraphics[width=0.8\textwidth]{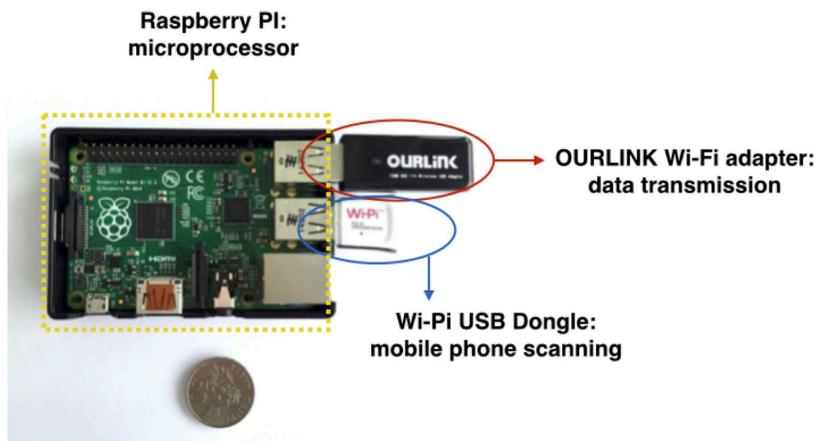}
\caption{The GN is a Raspberry PI connecting with Wi-Pi (white colour) and OURLINK (black colour) wireless interfaces.}
\label{fig_gn}
\end{figure}

We implement the GN (shown in Figure~\ref{fig_gn}) based on Raspberry PI, which connects with a Wi-Pi USB Dongle for probe request collection, and a ourLink Wi-Fi adapter so that the dataset of GN can be transmitted to the server through existing Wi-Fi network. Both of them work in 2.4GHz. The reason of using two wireless interfaces is that dual wireless interface system achieves probe requests collection while transmitting the datasets simultaneously. 

\begin{figure}[htb]
\centering
\includegraphics[width=0.8\textwidth]{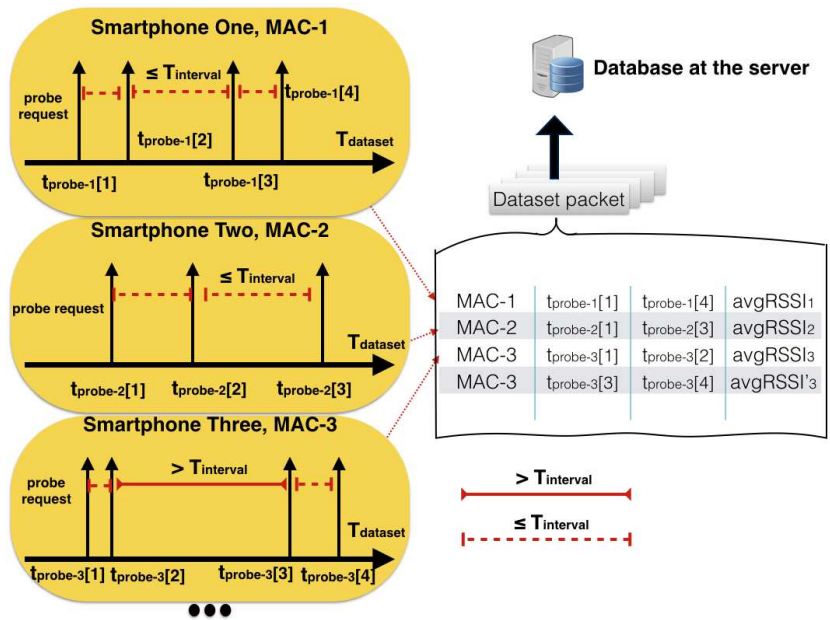}
\caption{PRI-DCS of \textit{SenseFlow}. In each $T_{dataset}$, MAC address is provided by the probe request from the smartphone. }
\label{fig_datastructure}
\end{figure}

Different smartphone models may have different probe request definitions. Generally, the probe request contains the Type of Frame, Duration, Source MAC address, BSSID, SEQ, and etc~\cite{freudiger2015talkative}.  
We propose a Probe Request Interval-based Data Collection Scheme (PRI-DCS) for \textit{SenseFlow}. The PRI-DCS is shown in Figure~\ref{fig_datastructure} and Algorithm~\ref{alg_DCA}. 
$T_{dataset}$ is the time to transmit the dataset packet. $t_{probe-\star}[\cdot]$ denotes the time of the probe request from the smartphone in $T_{dataset}$, e.g., $t_{probe-1}[2]$ is the second probe request from smartphone One. We define $T_{interval}$ as a time threshold to merge two probe requests of the smartphone. Specifically, if the time interval between two consecutive probe requests is smaller than $T_{interval}$, the smartphone is assumed to be not moving, and only one record that contains timestamp of the first probe request and the last one is kept by the PRI-DCS. Otherwise, both timestamps are kept in two independent records as the smartphone may leave and return to the coverage of the GN. Our aim to do so is to reduce the packet size to be uploaded. Finally, the MAC address of the smartphone, timestamps and average of RSSI of the probe requests are added to the dataset packet that is transmitted to the server for people tracking (details are presented in Section~\ref{hybrid}). 

\begin{algorithm}[t]
\caption{PRI-DCS Algorithm}
\label{alg_DCA}
\begin{algorithmic}[1]
\STATE{The timestamp of the latest probe request of smartphone $i$ is $t_{probe-i}$.}
\IF{A new probe request $x_{i}$ is received by GN $j$.}
\STATE{$x_{i} = x_{i} + 1$, the timestamp of $x_{i}$ is $t^{\prime}_{probe-i}$.}
\STATE{The connection time in $T_{dataset}$ is: $T(i, j) = t^{\prime}_{probe-i} - t_{probe-i}$.}
\IF{$T(i, j) < T_{interval}$}
\STATE{Update the latest timestamp of smartphone $i$ as $t^{\prime}_{probe-i}$ in $D_{j}(t)$.}
\ELSE
\STATE{Create a new record for smartphone $i$ with timestamp $t^{\prime}_{probe-i}$ in $D_{j}(t)$.}
\ENDIF
\IF{$System Clock == T_{dataset}$}
\STATE{Calculate average of RSSI for each MAC address.}
\STATE{GN $j$ uploads $D_{j}(t)$ to the server.}
\ENDIF
\ELSE
\STATE{The GN keeps listening.}
\ENDIF
\end{algorithmic}
\end{algorithm}

Most of people tracking systems are based on the probe requests received by the distributed sensor nodes from the smartphones. However, the sensor node at one location can receive the probe requests of a smartphone at an adjacent location due to an overlapping coverage area. As a result, those systems are not able to monitor accurate people density and track people flow since some of smartphones are captured at multiple locations at the same time, which we name \textit{probe request overhearing problem}. 
In \textit{SenseFlow}, we extend the PRI-DCS by utilising RSSI deviation at different locations to address this probe request overhearing problem. Although the RSSI of the probe request does not depict a precise location of the smartphone, it implies how far the smartphone is away from the GN since the RSSI measurements attenuate in distance with a power decay factor. Therefore, when multiple GNs receive the probe request from the same smartphone at the same time, the GN that is closest to the smartphone has the highest RSSI value. 
Specifically, in \textit{SenseFlow}, the GN calculates an average value of RSSI for each smartphone in $T_{dataset}$. The RSSI values are appended to the corresponding MAC address in the dataset packet (shown in Figure~\ref{fig_datastructure}). On the server, if any MAC address is captured in dataset packets from multiple GNs, we select the GN that has the maximum RSSI value as the location where the smartphone presents.

\subsection{\textit{SenseFlow} People Tracking Algorithm}
\label{hybrid}
In order to monitor people density and track people flow, a \textit{SenseFlow} People Tracking (SFPT) algorithm is proposed. Details are provided in Algorithm~\ref{alg_sensing}. Specifically, 
The SFPT algorithm amalgamates dataset packets to monitor people density and track people movement on the server. 
Based on the spatial-temporal dataset $D_{j}(t)$ of all GNs, the MAC address of smartphone $i$ with timestamp is extracted. To address the probe request overhearing problem, each MAC address only keeps one GN who has the highest RSSI at any time. Therefore, people density at any location is known by counting the unique MAC addresses that connect to the GN $j$ at that location. 

\begin{algorithm}[t]
\caption{SFPT Algorithm}
\label{alg_sensing}
\begin{algorithmic}[1]
\IF{$D_{j}(t)$ is received from the GN $j$}
\STATE{$D_{j}(t)$ is sorted by unique MAC address of the smartphone $i$.}
\FOR{$i \leq$ the total number of users}
\STATE{$[RSSI_{i},j^{\prime}]$ = max($RSSI_{i}$[1], $RSSI_{i}$[2], ..., $RSSI_{i}$[$j$], ...).}
\STATE{PeopleDensity[$j^{\prime}$] = PeopleDensity[$j^{\prime}$] + 1.}
\IF{${\overrightarrow{\mathcal{J}}}$ = \textbf{LCS}($\overrightarrow{\mathcal{X}_{i}}$, ${\overrightarrow{\mathcal{J}}}$)}
\STATE{The number of people in ${\overrightarrow{\mathcal{J}}}$ increases by 1.}
\ELSE
\STATE{Continue.}
\ENDIF
\ENDFOR
\ENDIF
\end{algorithmic}
\end{algorithm}

The trajectory of smartphone $i$ can be known by tracking the GNs that receive its probe requests according to time. Specifically, we formulate the trajectory as a sequence of GN IDs, which is denoted as $\overrightarrow{\mathcal{X}_{i}} = (x_{1}, x_{2}, ... x_{n})$, where $x_{n}$ is ID of the GN that has the strongest RSSI of probe request. 
Given a targeting trajectory ${\overrightarrow{\mathcal{J}}}$, we decide whether smartphone $i$ has ever travelled along ${\overrightarrow{\mathcal{J}}}$ based on the LCS of $\overrightarrow{\mathcal{X}_{i}}$. If all GNs in ${\overrightarrow{\mathcal{J}}}$ detect the probe requests from smartphone $i$, the LCS of $\overrightarrow{\mathcal{X}_{i}}$ and ${\overrightarrow{\mathcal{J}}}$ is ${\overrightarrow{\mathcal{J}}}$. The number of people moving along ${\overrightarrow{\mathcal{J}}}$ is obtained by counting the users who fulfil ${\overrightarrow{\mathcal{J}}} = LCS({\overrightarrow{\mathcal{J}}},\overrightarrow{\mathcal{X}_{i}})$. 

\begin{figure}[htb]
\centering
\includegraphics[width=0.8\textwidth]{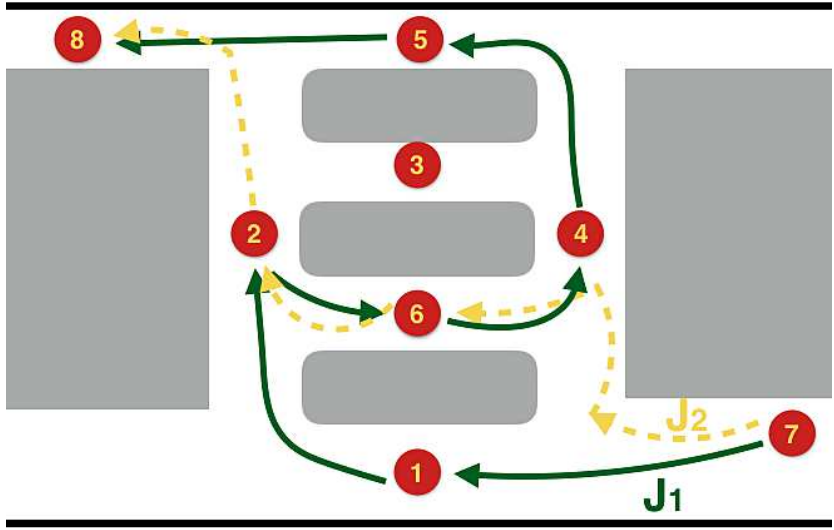}
\caption{An example for trajectory recognition: eight GNs are deployed along the walking path in an area.}
\label{fig_pathexample}
\end{figure}

For trajectory recognition, as an example shown in Fig~\ref{fig_pathexample}, we have eight GNs along the walking path in an area. From GN (7) to (8), we consider two targeting trajectories, ${\overrightarrow{\mathcal{J}_{1}}} = (7,1,2,6,4,5,8)$ and ${\overrightarrow{\mathcal{J}_{2}}} = (7,4,6,2,8)$. Assume the system detects a smartphone moving from GN (7) to (8), and the trajectories is given by $\overrightarrow{\mathcal{X}_{1}} = (7,1,6,5,8)$. Then, we have $LCS({\overrightarrow{\mathcal{J}_{1}}},\overrightarrow{\mathcal{X}_{1}}) = (7,1,6,5,8)$ and $LCS({\overrightarrow{\mathcal{J}_{2}}},\overrightarrow{\mathcal{X}_{1}}) = (7,6,8)$. Therefore, by using LCS, the system recognises that $\overrightarrow{\mathcal{X}_{1}}$ travels along ${\overrightarrow{\mathcal{J}_{1}}}$, not ${\overrightarrow{\mathcal{J}_{2}}}$ since $LCS({\overrightarrow{\mathcal{J}_{1}}},\overrightarrow{\mathcal{X}_{1}})$ has more nodes than $LCS({\overrightarrow{\mathcal{J}_{2}}},\overrightarrow{\mathcal{X}_{1}})$. 
Additionally, the more GNs detect the smartphone, the more accurate trajectory recognition \textit{SenseFlow} achieves. Consider an extreme case that only three GNs detect a smartphone, $\overrightarrow{\mathcal{X}_{2}} = (7,3,8)$. Then, we have $LCS({\overrightarrow{\mathcal{J}_{1}}},\overrightarrow{\mathcal{X}_{2}}) = (7,8)$ and $LCS({\overrightarrow{\mathcal{J}_{2}}},\overrightarrow{\mathcal{X}_{1}}) = (7,8)$. The trajectory of the smartphone is not be able to be tracked since it is hardly detected by GNs. 

\section{Experiments On Testbed and Evaluation}
\label{experiment}
In this section, we first present the experiments to show the effect of probe request transmitting interval on $SenseFlow$. Then, the extensive experiments for monitoring people density are conducted in four classrooms and one laboratory on SUTD university campus, and a crowded public area in Singapore. Next, we measure how people walking behaviour effects the probe request detection under different smartphone's operational modes. We evaluate the performance of $SenseFlow$ for tracking people flow on our testbed in the university.

\subsection{Probe Request Interval Measurement}
\label{expBeacon}
Generally, on any smartphone, probe requests are transmitted in bursts, the interval of which depends on the OS and Wi-Fi chipset driver and varies greatly according to status of the Wi-Fi connection and screen mode~\cite{demir2013wi,JamesCisco}. Some of smartphones may not be captured since the probe request interval, $(t^{\prime}_{probe-i} - t_{probe-i})$ (in \textit{Alg.}~\ref{alg_DCA}) is longer than the time when the people move across the sensing range of GN, which degrades the tracking accuracy of the system. To understand how smartphones affect the performance, we evaluate the system with four operational modes, (Wi-Fi registered, screen on), (Wi-Fi non-registered, screen on), (Wi-Fi registered, screen off), and (Wi-Fi non-registered, screen off). Specifically, ``Wi-Fi registered" implies that the phone is authorised to connect to a Wi-Fi network. 

In this experiment, we choose three typical smartphone models, two iOS phones (one iPhone 4 and one iPhone 4S), five Android phones (one Samsung Galaxy Nexus, one ASUS MeMo pad, three Sony Xperia Z5) and one Windows phone (Nokia Lumia 520). 
Table~\ref{tb_interval} presents average probe request interval of smartphones in different Wi-Fi and screen conditions. Specifically, in screen off and Wi-Fi non-registered mode, the smartphones increase the probe request interval to conserve battery power, comparing to the mode of screen on. Moreover, iOS phone and Windows phone have a long interval around 1200 seconds when the smartphone has connected to Wi-Fi network. However, Android phone still keeps a short interval of 2.11 and 2.15 seconds. The different probe request interval is caused by a differentiated energy-saving design of smartphones in Wi-Fi registered mode. 

\begin{table}[htb]
    \centering
    \caption{Average probe request interval of smartphones in different Wi-Fi and screen mode.}
    \begin{tabular} {p{2.9cm}|p{1.3cm} p{1.3cm}|p{1.3cm} p{1.2cm}} \hline
        \bf{Smartphones} & \multicolumn{2}{c|}{\bf{non-registered Wi-Fi}} & \multicolumn{2}{c}{\bf{registered Wi-Fi}} \\ \hline
	\bf{} & \bf{screen on} & \bf{screen off} & \bf{screen on} & \bf{screen off}\\ \hline 
        	iOS &  70.6s & 109.8s &  1200.8s & 1204.4s\\ 
	Android &  0.8s & 1s & 2.11s &  2.15s \\ 
        	Windows &  10.9s & 13.9s & 1200.8s & 1204.4s\\ \hline 
   \end{tabular}
\label{tb_interval}
\end{table}

\subsection{$T_{dataset}$ and $T_{interval}$ Characterisation}
\label{expDatasize}
The GN in \textit{SenseFlow} transmits the dataset to the public network wirelessly. A practical question is how much data traffic will the GN generate everyday? This issue is crucial when the data is forwarded to the server via cellular network since more data traffic causes higher data bill. Therefore, we next study the impact of $T_{dataset}$ and $T_{interval}$ on the daily data size. In this experiment, we deploy 12 GNs with different $T_{dataset}$ and $T_{interval}$ configuration in a lab on the SUTD University campus. We run the experiment for one day (1440 minutes), and analyse the total data size collected from the GN. 

\begin{figure}[htb]
\centering
\includegraphics[width=0.8\textwidth]{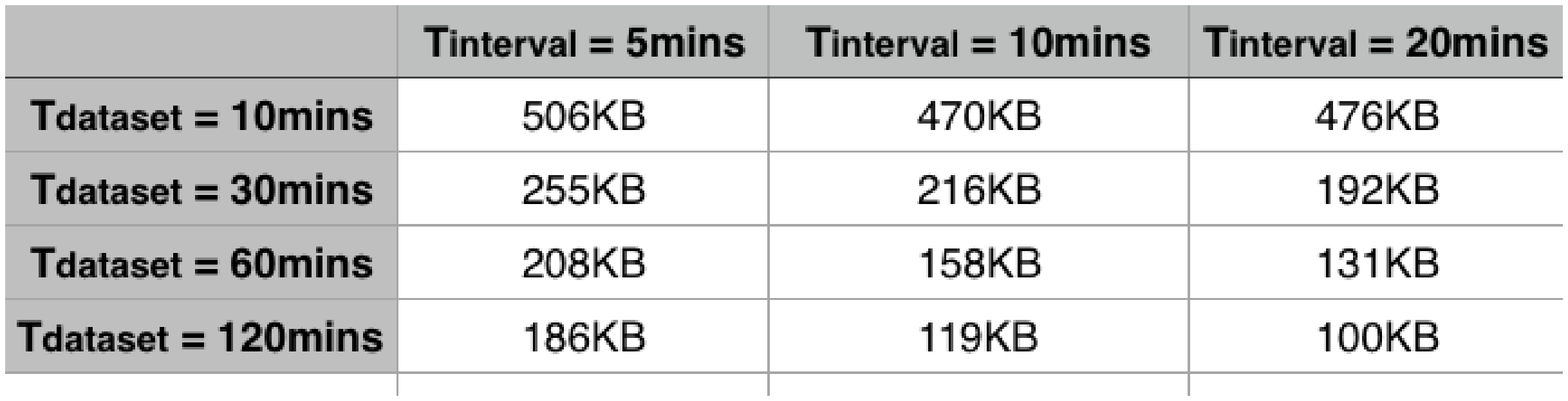} 
\caption{Data traffic with different $T_{dataset}$ and $T_{interval}$.} 
\label{fig_expData}
\end{figure}

The experimental results are shown by Figure~\ref{fig_expData}. The maximum data traffic is 506KB when  $T_{dataset}$ is 10mins and $T_{interval}$ is 5mins. With an increase of $T_{dataset}$, the GN uploads data in a long transmission interval where the unique MAC address is merged and the number of records is reduced significantly. For example, given $T_{interval}$=5mins and $T_{dataset}$=120mins, the daily data size is reduced to 186KB. Moreover, Figure~\ref{fig_expData} shows that increasing $T_{interval}$ also reduces data traffic. The reason is the records of the MAC address that fulfils $(t^{\prime}_{probe-i} - t_{probe-i}) < T_{interval}$ are merged to one record. 

\subsection{People Density Monitoring}
\label{expDensity}
To monitor people density in a public area, we deploy \textit{SenseFlow} in three applications: one lab room in the university to observe office hours of researchers, a crowded area in city to learn people density in public, and four closely located classrooms on the SUTD University campus to study the number of students in the classrooms. 

\subsubsection{Laboratory In University}
\begin{figure}[htb]
\centering
\includegraphics[width=0.5\textwidth]{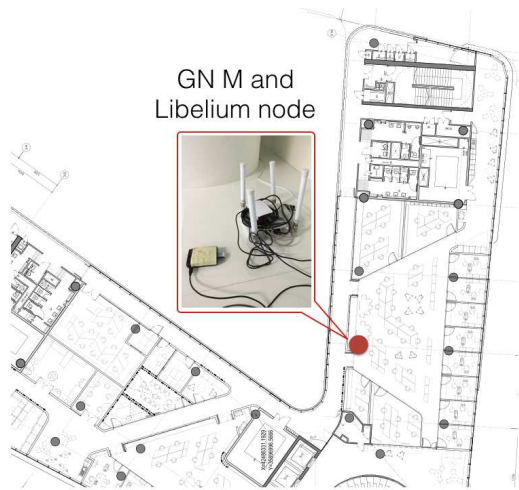}
\caption{GN M and Libelium device are deployed at the same location in a lab in the university.}
\label{fig_office}
\end{figure}

We deploy one GN and one off-the-shelf Meshlium model, produced by Libelium, in a lab room on campus for 7 days X 24 hours continuous people tracking, as shown in Figure~\ref{fig_office}. They are put at the same location. Furthermore, we personally visit the lab, and record the number of people by head count from 10AM to 6PM every weekday as the ground truth of this experiment. 

\begin{figure*}[htb]
\begin{center}
\begin{tabular}{cc}
\includegraphics[width=0.55\textwidth]{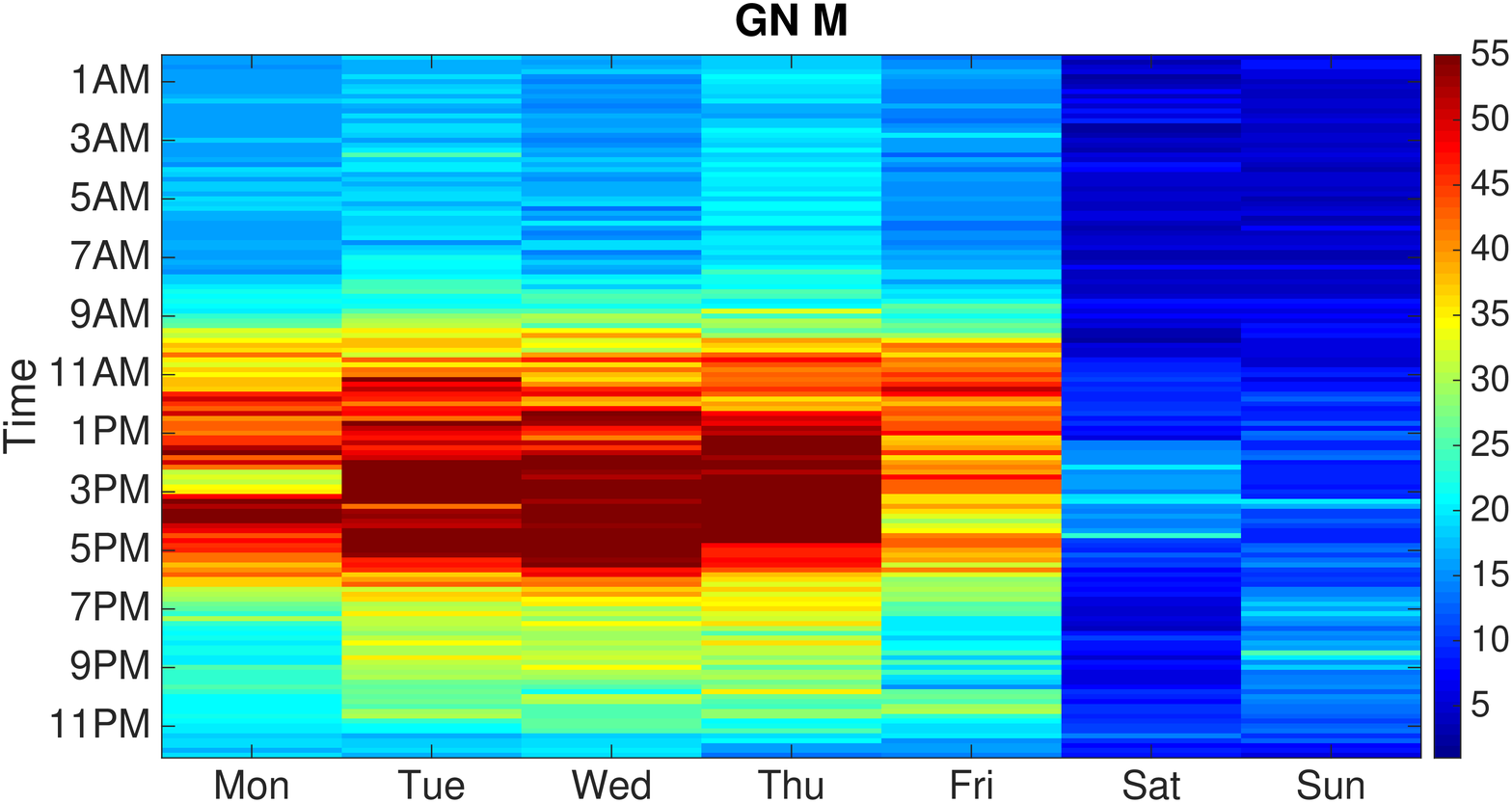} & \includegraphics[width=0.55\textwidth]{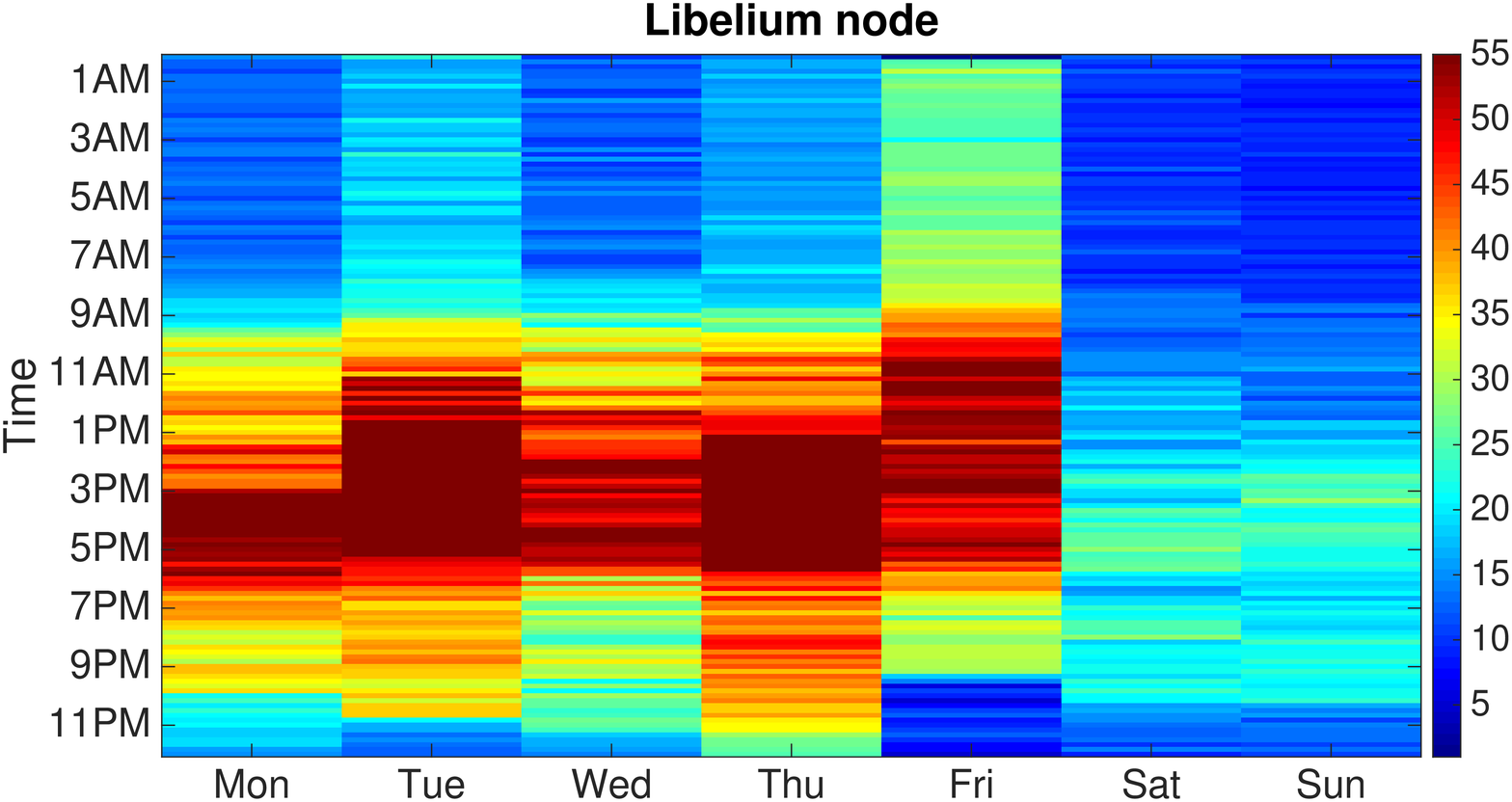} 
\\ (a) Smartphones detected by GN M & (b) Smartphones detected by Libelium node
\end{tabular}
\\ \includegraphics[width=0.6\textwidth]{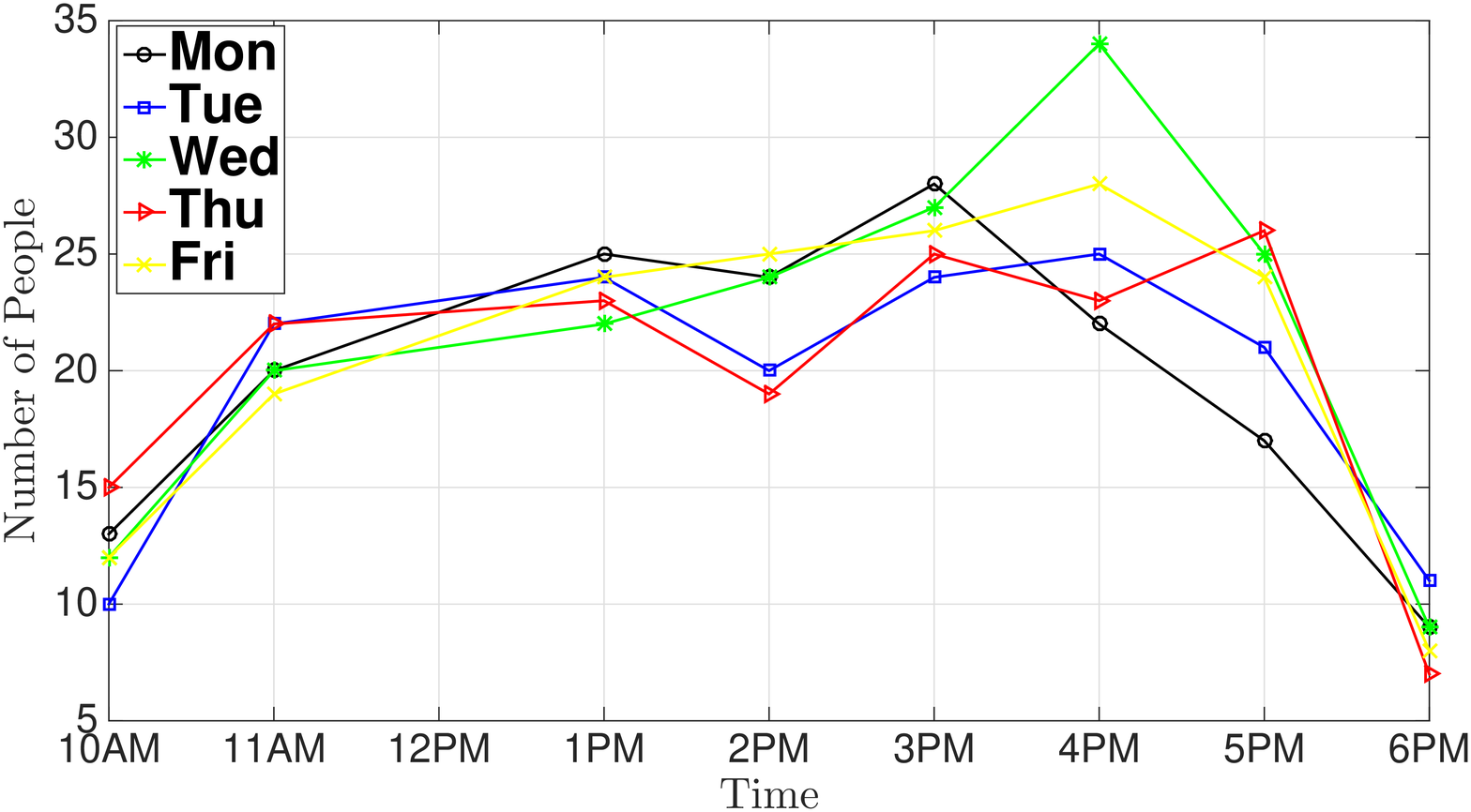} 
\\ (c) Number of people at specific time recorded by head count
\end{center}
\caption{People density monitoring in a lab in the university from Monday to Sunday (from 4th Jan 2016 to 10th Jan 2016).}
\label{fig_oncampus}
\end{figure*}

Figure~\ref{fig_oncampus}(a) and (b) show the number of unique smartphones detected by the GN M and Libelium node. Generally, the amount of smartphones detected by Libelium node is larger than the detections of GN M for around 10 phones. The reason is Libelium node has higher packet receiving sensitivity and larger signal coverage. Therefore, Libelium node scans more smartphones from other rooms. From Monday to Friday, both of the nodes detect more people in the daytime, from 8AM to 5PM, than the time before dawn and midnight. Specifically, the amount of smartphones detected by our GN is closer to the number of people recorded by head count during daily office hours, which is shown in Figure~\ref{fig_oncampus}(c). 
On weekends, people density decreases significantly. Less than 20 smartphones are detected in the lab. 

\subsubsection{Crowded Public Area In City}
\begin{figure}[htb]
\begin{center}
\begin{tabular}{cc}
\includegraphics[width=0.4\textwidth]{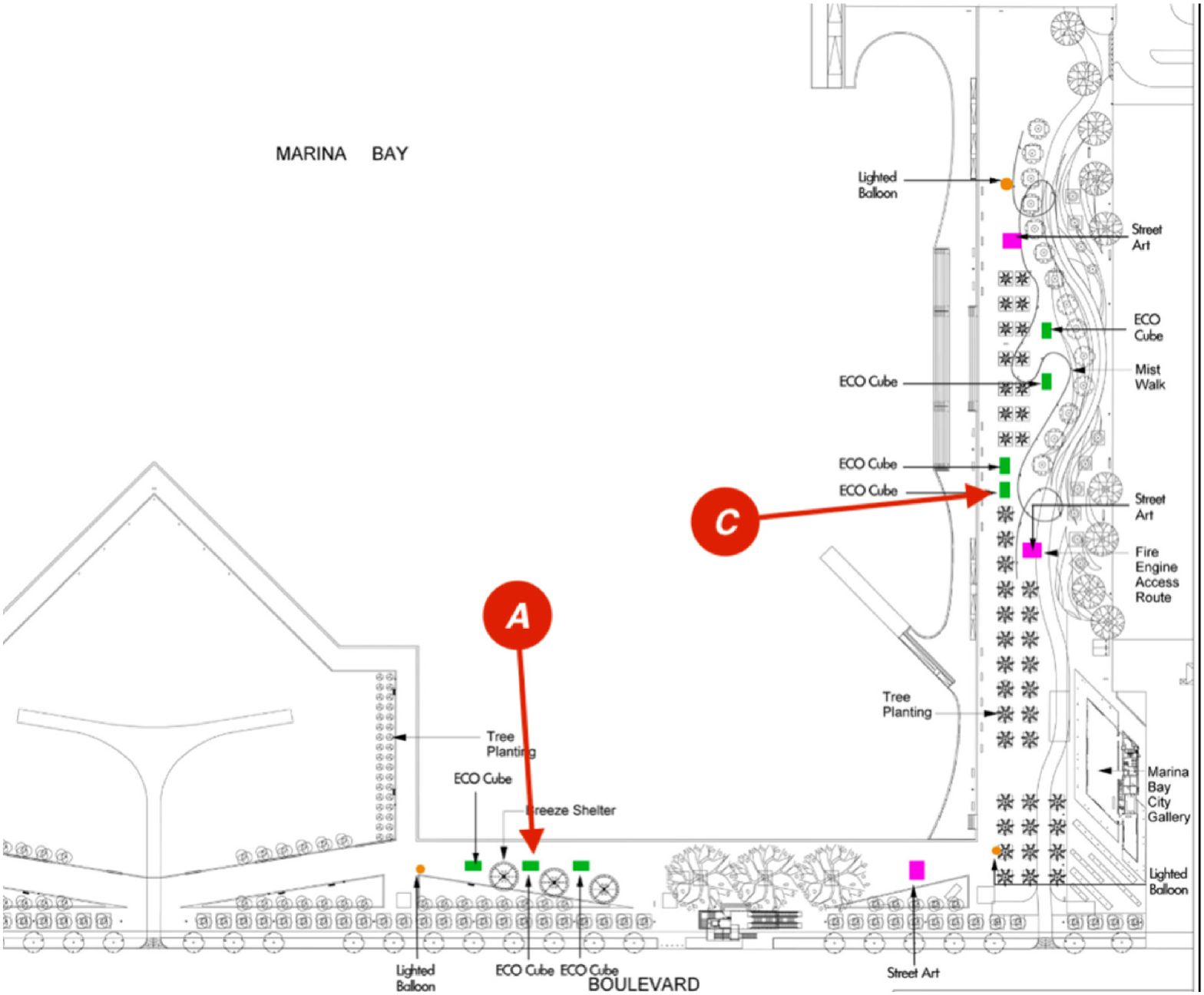} & \includegraphics[width=0.6\textwidth]{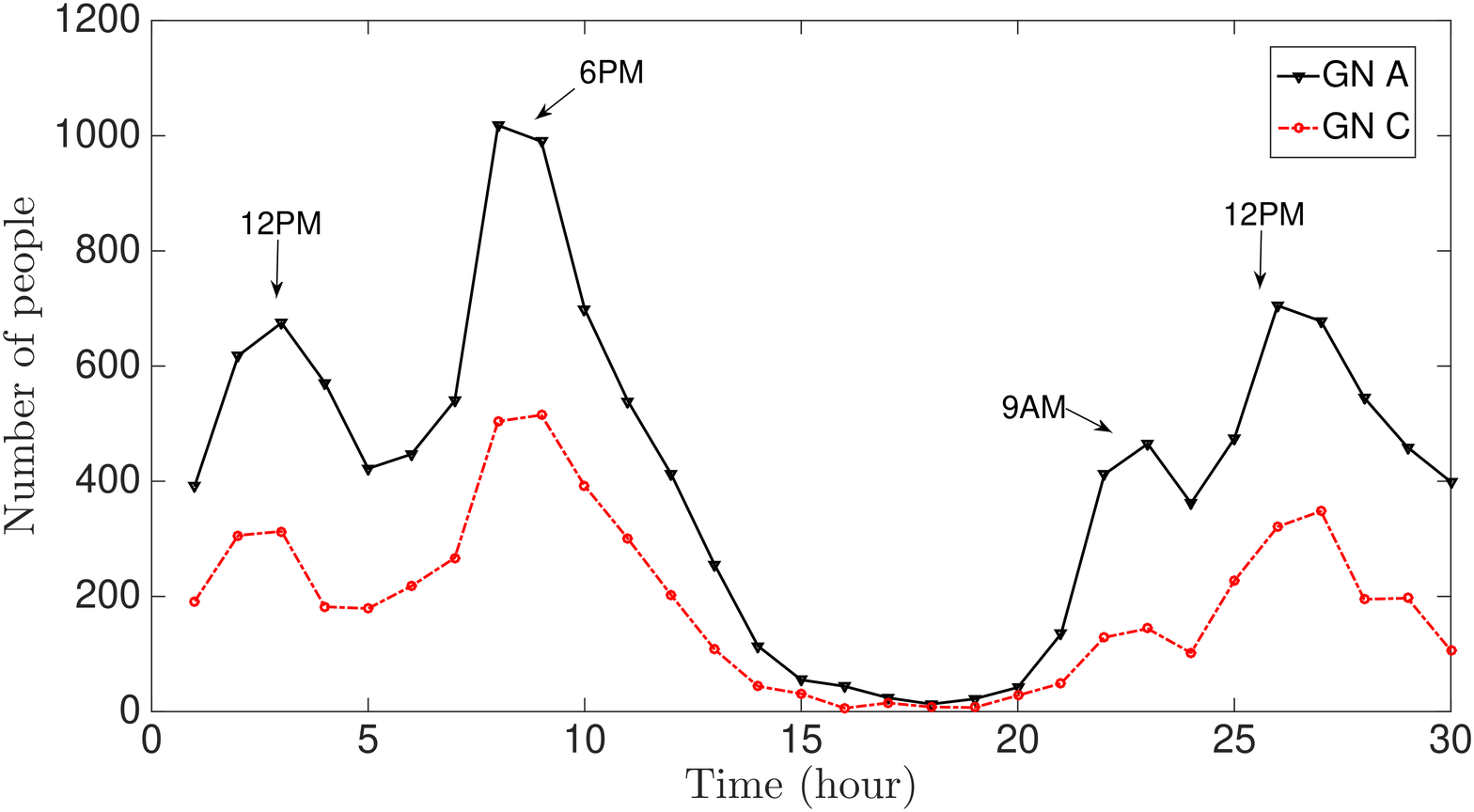} 
\\ (a) GNs A and C are deployed \\ in a public area. & (b) The number of unique smartphones\\ 
& connects to GN A or C in the monitoring area.
\end{tabular}
\\ \includegraphics[width=0.7\textwidth]{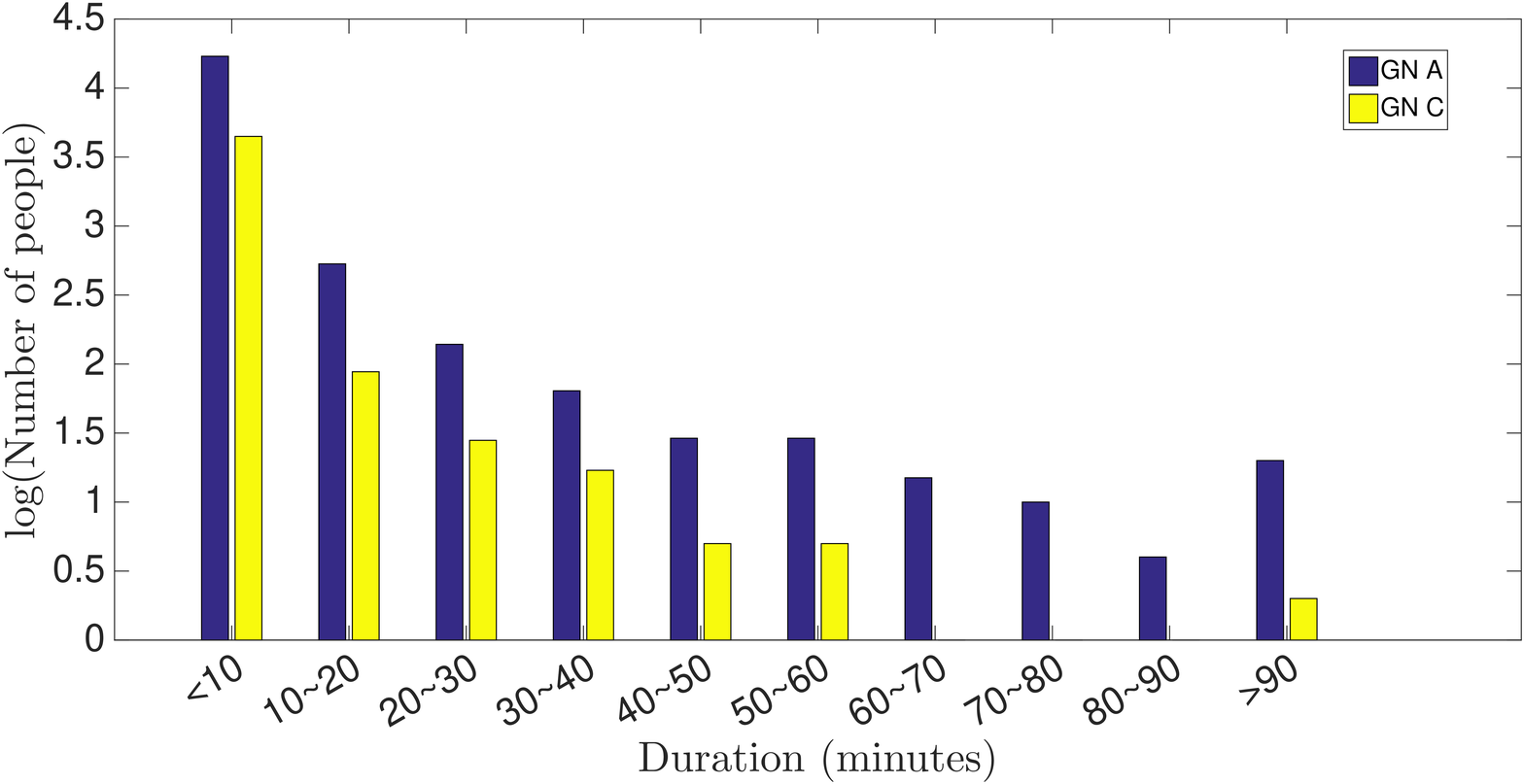} 
\\ (c) The number of devices connecting to GN A or C in a specific duration.
\end{center}
\caption{People density monitoring in a crowded city area in Singapore (from 11PM 28 Oct, 2014 to 4PM 29 Oct, 2014).}
\label{fig_offcampus}
\end{figure}

To test system scalability, two GNs, GN A and GN C, are placed along the walking path in a crowded city area in Singapore, as shown in Figure~\ref{fig_offcampus}(a). The experiment was carried out from 11PM 28 Oct, 2014 to 4PM 29 Oct, 2014 (30 hours in total). The people counting performance is shown in Figure~\ref{fig_offcampus}(b). It can be observed that there are three peaks of people quantity at 9AM, 12PM and 6PM in one day. The results indicates that they are rush hours and many people go through the public area. Moreover, more smartphones connecting to GN A than GN C indicates that the location of GN A is more popular than the one of GN C in the monitoring area. This result can be used for guidance of city planning and promotion of coordinated development of the public area. 

Based on the SFPT algorithm in Section~\ref{hybrid}, how long each people stays in the monitoring area can be known. Due to a large variance on the people, Figure~\ref{fig_offcampus}(c) presents the number of devices connecting to GN A or GN C during one day in $log$ scale. Specifically, the amount of devices whose connection time to GN A and GN C is less than 10 minutes is about 17008 and 4462. A few smartphones connects to the GNs for more than 20 minutes. This indicates that most of people could just pass by the GNs. Moreover, there are only 20 and 2 smartphones connecting with GN A or C for longer than 90 minutes. Those people could be the staff who works in the nearby shops. The significance of this experiment and result is that the building planner can make a more efficient plan based on information of the people quantity and flow movement. 

\subsubsection{Classrooms In University}
\begin{figure}[htb]
\begin{center}
\begin{tabular}{cc}
\includegraphics[width=0.6\textwidth]{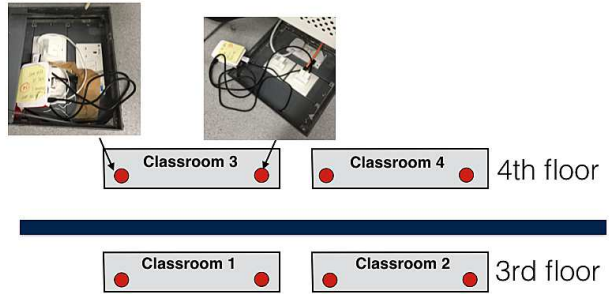} 
\\ (a) Eight GNs are deployed in four adjacent classrooms. Each classroom has two GNs. \\
\includegraphics[width=0.8\textwidth]{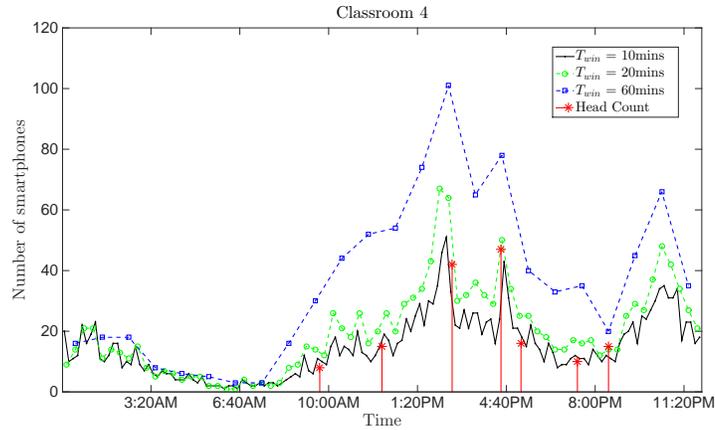} 
\\ (b) The number of smartphones detected by \textit{SenseFlow} compared with ground truth. 
\end{tabular}
\end{center}
\caption{The number of smartphones in the Classroom Four on 18th Nov 2015 (from 0:00:00AM to 11:59:59PM).}
\label{fig_class}
\end{figure}

To monitor people density, eight GNs are deployed in four adjacent classrooms. Classrooms One and Two are in the 3rd floor of the building, and classrooms Three and Four are right over One and Two in the 4th floor. The GNs are closely located, as shown in Figure~\ref{fig_class}(a). 

Figure~\ref{fig_class}(b) presents the number of unique smartphones in the Classroom Four for one day (18th Nov 2015) when the data analysing time window, $T_{win}$, is 10mins, 20mins, or 60mins. Namely, we calculate the number of unique smartphones every $T_{win}$. 
It is observed that the number of smartphones from daytime to midnight are generally more than the ones before dawn.  
In this experiment, we personally visit the classrooms at different time to count the number of students. We use these recorded head counts to compare with \textit{SenseFlow} when $T_{win}$ = 10mins since the ground truth is based on the discrete time point. 
Generally, the difference between \textit{SenseFlow} and ground truth is less than 2. It is also observed that \textit{SenseFlow} with $T_{win}$ = 10mins undercounts 6, 7 and 3 people comparing to the ground truth at 2:00PM, 4:30PM, and 8:30PM. A possible reason is that there is an event in the classroom where some of the students switch off their mobile devices. 
Additionally, it is observed from $T_{win}$ = 60mins that around 100 different smartphones are in the classroom from 1PM to 2PM, which indicates the peak hour in the classroom Four. 

\begin{figure}[htb]
\centering
\includegraphics[width=0.8\textwidth]{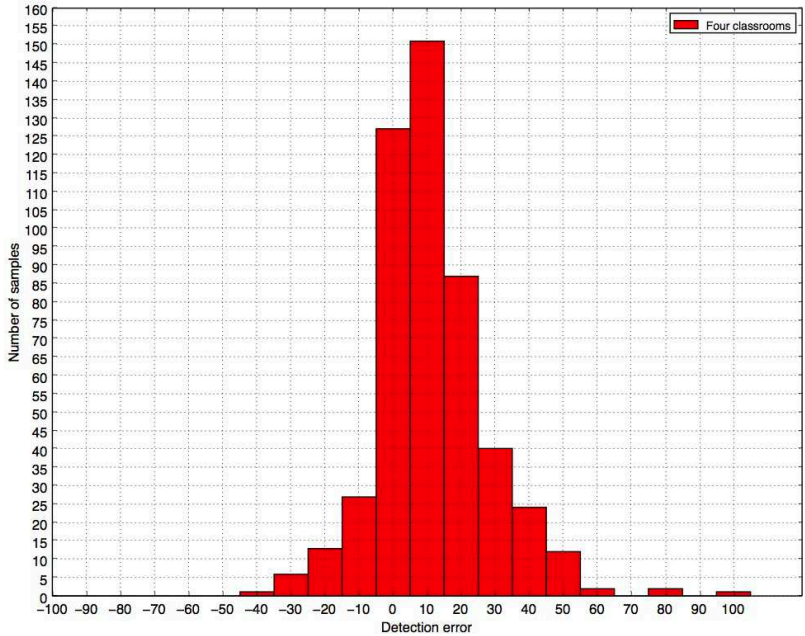}
\caption{The detection error of \textit{SenseFlow} in the four classrooms. The figure shows the performance for 33 days.}
\label{fig_detectionError}
\end{figure}

The data of \textit{SenseFlow} is collected over 33 days in total, and we record 130 samples for the ground truth. To compare the number of mobile phones detected by \textit{SenseFlow} with the actual number of people in classrooms, we define system detection error at a specific time as follows.  
\begin{equation}
\text{Detection error} = \frac{(\text{number of phones detected by \textit{SenseFlow}}) - (\text{ground truth})}{\text{ground truth}} 
\end{equation}
where the data of \textit{SenseFlow} is based on $T_{win}$ = 10mins, which is comparable to the ground truth. Moreover, we select the data to compare given the non-zero ground truth value. 
Therefore, we have three possible results of detection error: 
\begin{description}
\item [the detection error $>$ 0] the number of phones detected by \textit{SenseFlow} is more than the number of people in the four classrooms. 
\item [the detection error $=$ 0] the number of phones detected by \textit{SenseFlow} is the same as the number of people in the four classrooms. 
\item [the detection error $<$ 0] the number of phones detected by \textit{SenseFlow} is lower than the number of people in the four classrooms. 
\end{description}
Figure~\ref{fig_detectionError} presents the histogram of detection error over the four classrooms. As observed, 127 samples are as accurate as the ground truth, 151 samples have 10\% detection error, and 87 samples have 20\% detection error. 

\subsection{Human Walking Behaviour Effect}
\label{expSpeed}
\begin{figure*}[htb]
\begin{center}
\begin{tabular}{cc}
\includegraphics[width=0.55\textwidth]{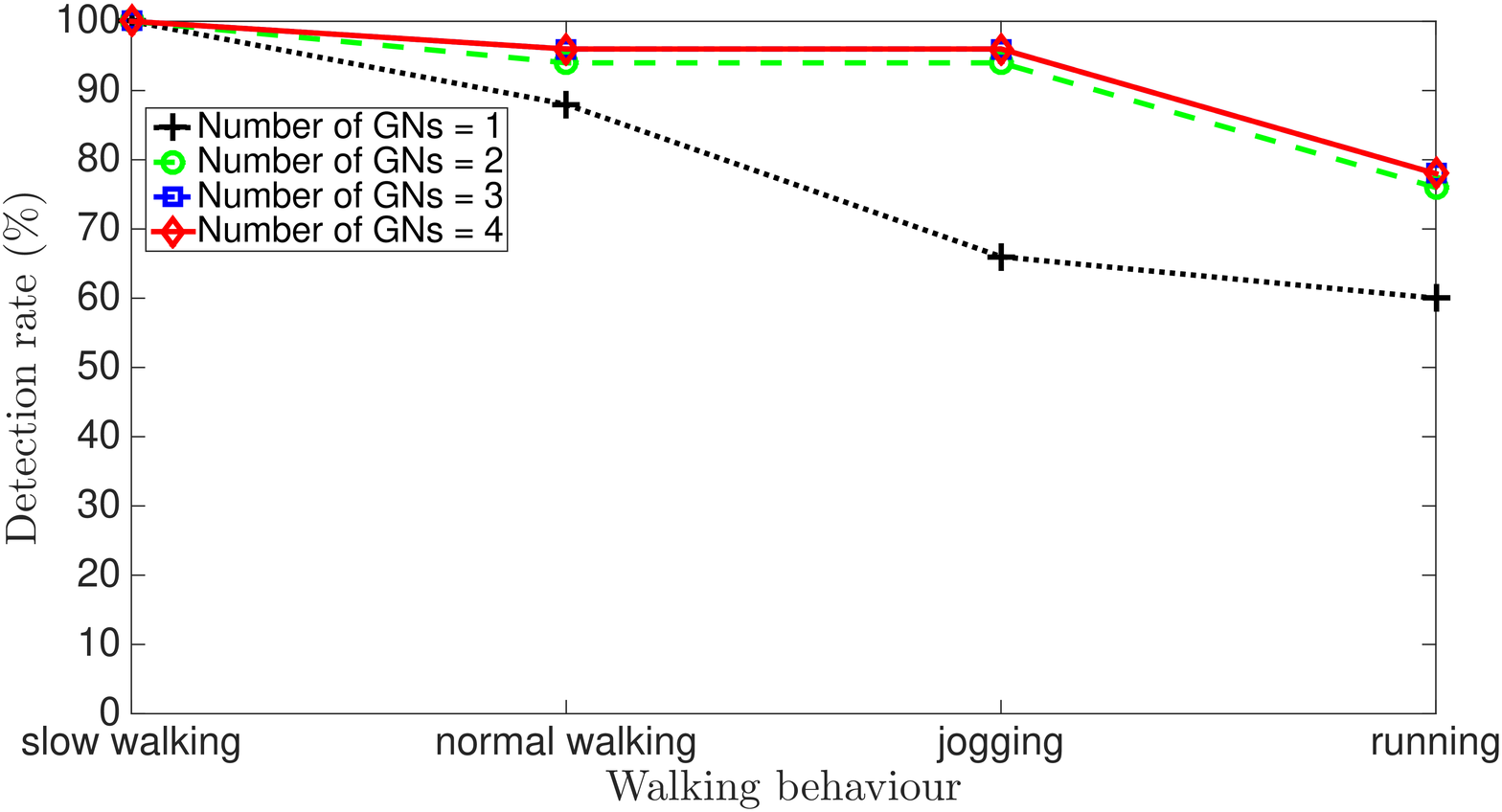} & \includegraphics[width=0.55\textwidth]{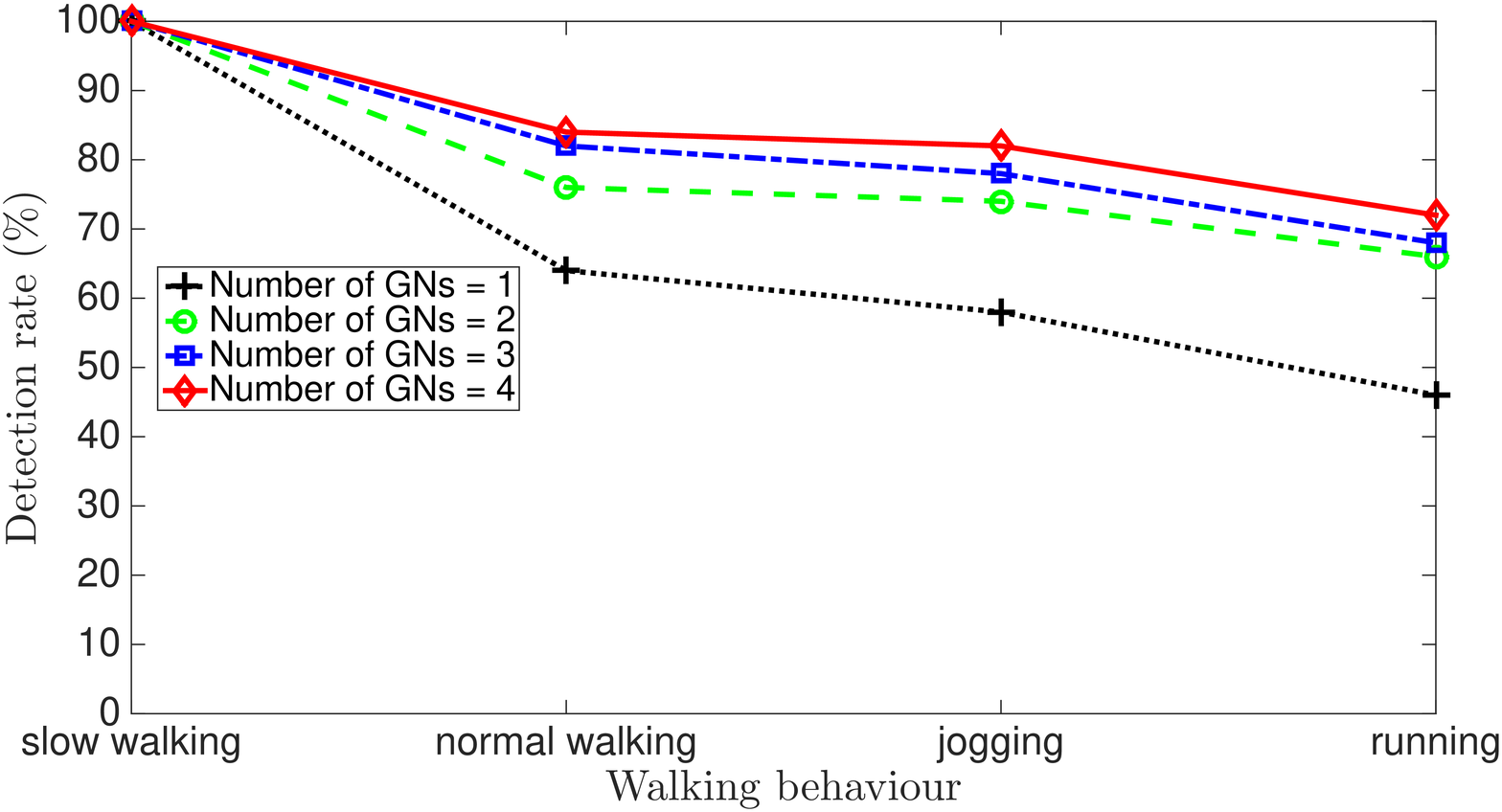} 
\\ (a)(screen on, Wi-Fi non-registered) & (b) (screen off, Wi-Fi non-registered) 
\\ \includegraphics[width=0.55\textwidth]{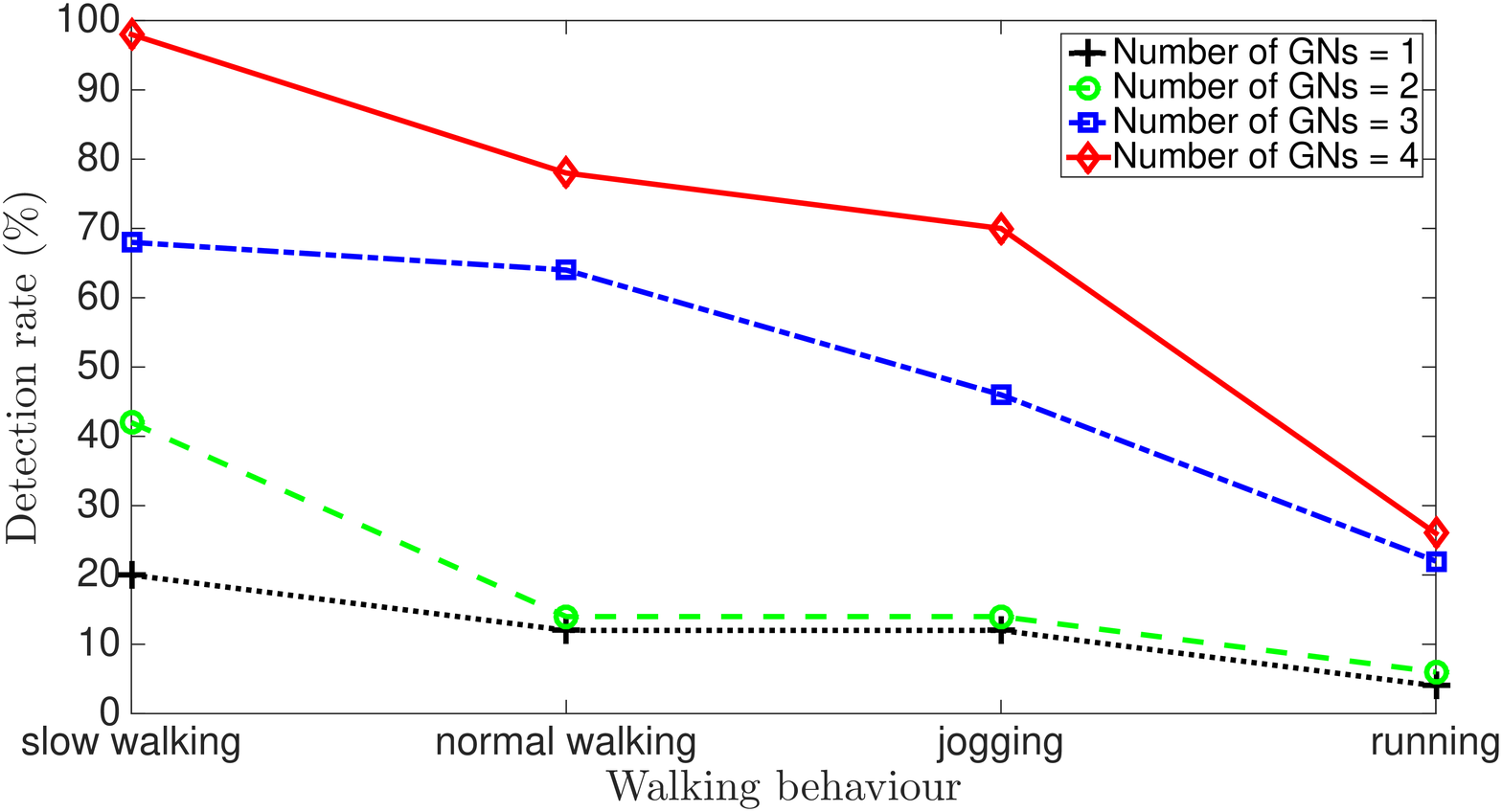} & \includegraphics[width=0.55\textwidth]{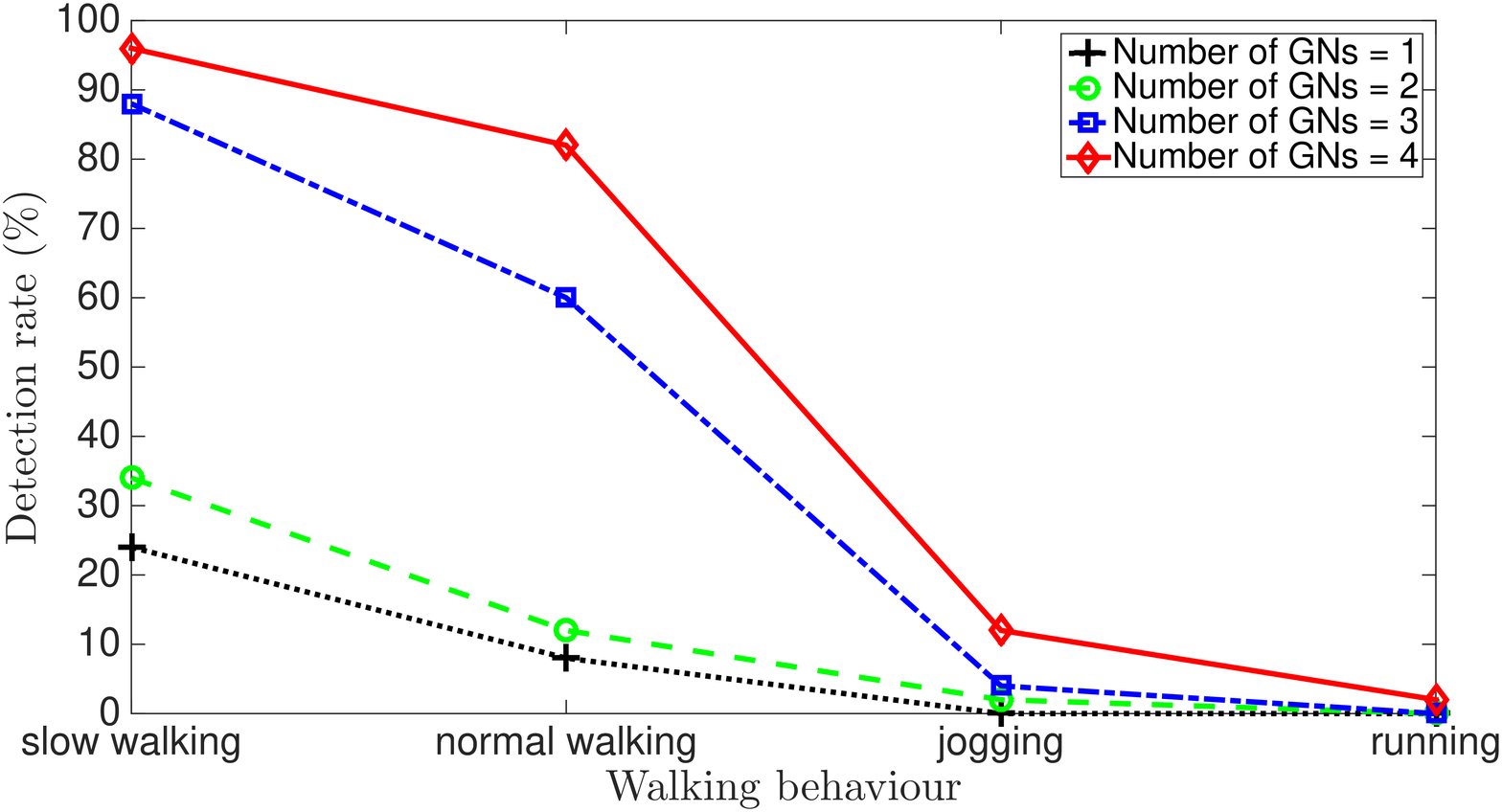}
\\ (c) (screen on, Wi-Fi registered) & (d) (screen off, Wi-Fi registered) 
\end{tabular}
\end{center}
\caption{Detection rate of the smartphones under different human walking behaviours and operational modes.}
\label{fig_expspeed}
\end{figure*}
From Table~\ref{tb_interval}, it is observed that a longer contact time between the GN and smartphone can increase the probe request receiving probability. Namely, the slower the people move, the higher chance that the GN can capture the probe requests from the smartphone. Therefore, we further measure the effect of people walking behaviour on \textit{SenseFlow}. 

In this experiment, the GNs are deployed in a monitoring area to capture the user's presence. One person carrying five smartphones, i.e., two iOS phones, two Android phones, and one Windows phone, goes through the area with different walking speed. We employ four human walking behaviours, \textit{slow walking}, \textit{normal walking}, \textit{jogging} and \textit{running} with referred speeds of 1.25, 2.25, 2.6, and 4.5\textit{m/s}, respectively. The four operational modes (described in Section~\ref{expBeacon}) are also considered. In each experiment, we increase the number of GNs from 1 to 4. For each operation mode and walking behaviour, we repeat the experiment for ten times, therefore, detection rate equals to the average number of smartphones that are detected by any one of GNs. 

Figure~\ref{fig_expspeed} shows the performance of detection rate under different operational modes and walking behaviours. Generally, the detection rate grows with an increase in the number of GNs. However, with an increase of walking speed, the detection rate of smartphones decreases. Because the probe request is unable to be detected when its interval is longer than the contact time of smartphone and the GN. 
Moreover, \textit{SenseFlow} achieves the highest detection rate when the smartphone is in (screen on, Wi-Fi non-registered) mode. The reason is the smartphone's energy-saving design prolongs the probe request interval when the screen is turned off or Wi-Fi is connected.

\subsection{People Flow Tracking Experiments}
\label{expTracking}
We evaluate the performance of \textit{SenseFlow} by deploying a proof of concept testbed on the SUTD University campus. The testbed consists of fourteen GNs which are evenly deployed at seven locations in four buildings of the university, and each location contains two GNs where are one meter away from each other. The location sequence is from 1 to 7. Three people move from location 1 to 7 at normal walking speed and each of them carries one smartphone. The smartphones in this experiment are one iOS, one Android and one Windows phone. We choose two targeting trajectories from the same starting point (location 1) to the same destination (location 7), $\overrightarrow{\mathcal{J}_{1}} = (1, 2, 3, 4, 5, 6, 7)$ and $\overrightarrow{\mathcal{J}_{2}} = (1, 2, 5, 6, 4, 3, 7)$. The starting and ending time are denoted as $t_{1}$ and $t_{7}$ respectively. 

Given the trajectory of smartphone $i$ is $\overrightarrow{\mathcal{X}_{i}}$, the length of $\overrightarrow{\mathcal{X}_{i}}$ is presented by $\ell(\overrightarrow{\mathcal{X}_{i}})$. Moreover, the number of locations successfully detects the smartphone is $\ell(\overrightarrow{\mathcal{X}^{\prime}_{i}})$. Therefore, the tracking accuracy can be given by 
\begin{equation}
\delta = \frac{\ell(\overrightarrow{\mathcal{X}^{\prime}_{i}})}{\ell(\overrightarrow{\mathcal{X}_{i}})}
\end{equation}

\begin{figure}[htb]
\centering
\includegraphics[width=0.8\textwidth]{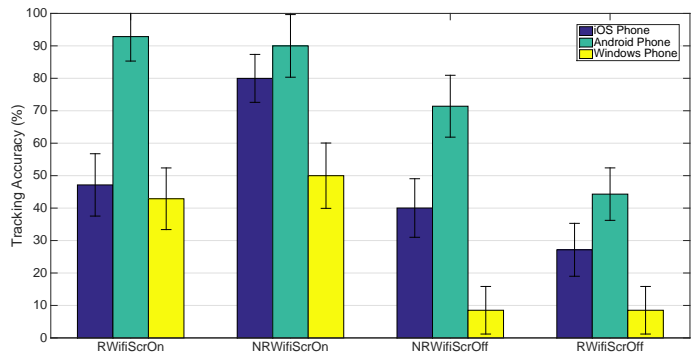}
\caption{Tracking accuracy of iOS, Android and Windows phone in four modes, (screen on, Wi-Fi registered), (screen on, Wi-Fi non-registered), (screen off, Wi-Fi non-registered), and (screen off, Wi-Fi registered). The confidence interval is based on 10 experiments. }
\label{fig_exp_trackingaccuracy}
\end{figure}

Figure \ref{fig_exp_trackingaccuracy} shows the performance of \textit{SenseFlow} with different smartphone models and operational modes. In RWifiScrOn (screen on, Wi-Fi registered), Android phone achieves the highest $\delta$, about 92.9\%. iOS and Windows phone achieve $\delta = 47.2\%$ and $\delta = 42.9\%$ on average. 
In NRWifiScrOn (screen on, Wi-Fi non-registered), the average $\delta$ of Android, iOS and Windows phone is 90\%, 80\% and 50\%, respectively. The tracking accuracy of iOS and Windows phone in this case is higher than the one in RWifiScrOn. The reason is that when the Wi-Fi of the smartphones is not connected, they broadcast probe requests more frequently in order to search the Wi-Fi connection. 
NRWifiScrOff (screen off, Wi-Fi non-registered) and RWifiScrOff (screen off, Wi-Fi registered) present the walking trajectories tracked by \textit{SenseFlow} when the screen of smartphone is off. Specifically, the $\delta$ of Android phone is 71.4\% and iOS is 40\% in NRWifiScrOff. The $\delta$ of Android and iOS go down to 44.3\% and 27.2\% in RWifiScrOff. This is because the smartphone increases the probe request interval when Wi-Fi is connected. Moreover, in both experiments, the trajectory of Windows phone is not tracked since its wireless transceiver is turned off when its screen is off. 

The recognition of the two targeting trajectories, $\overrightarrow{\mathcal{J}_{1}}$ and $\overrightarrow{\mathcal{J}_{2}}$, depends on $\overrightarrow{\mathcal{J}_{1}}^{\star} = (3, 4, 5, 6)$ and $\overrightarrow{\mathcal{J}_{2}}^{\star} = (5, 6, 4, 3)$. Furthermore, since the locations 5 and 6 are in the same order in $\overrightarrow{\mathcal{J}_{1}}$ and $\overrightarrow{\mathcal{J}_{2}}$, for simplicity, we denote them as one combined location (5, 6), then we have $\overrightarrow{\mathcal{J}_{1}}^{\star} = (3, 4, (5, 6))$ and $\overrightarrow{\mathcal{J}_{2}}^{\star} = ((5, 6), 4, 3)$. 

To evaluate the trajectories recognition of $SenseFlow$, we define successful trajectory recognition rate which is a ratio of number of phones on the targeting trajectory and the total number of phones. In this experiment, five people walk from location 1 to location 7 along the trajectories $\overrightarrow{\mathcal{J}_{1}}$ and $\overrightarrow{\mathcal{J}_{2}}$, and each of them carries one smartphone. The types of smartphone we use include two android phones, two iPhones and one Windows phone. 
We find that the highest successful recognition rate of the two targeting trajectories, $\overrightarrow{\mathcal{J}_{1}}^{\star}$ and $\overrightarrow{\mathcal{J}_{2}}^{\star}$, is 100\% when the phones are in the mode, (screen on, Wi-Fi non-registered). The reason is the smartphone transmits probe requests frequently in this operational mode. The GNs at the locations in $\overrightarrow{\mathcal{J}_{1}}^{\star}$ and $\overrightarrow{\mathcal{J}_{2}}^{\star}$ detect the smartphone successfully. 
In (screen off, Wi-Fi non-registered) mode, \textit{SenseFlow} recognises 40\% of trajectories on $\overrightarrow{\mathcal{J}_{2}}$, and 60\% of trajectories on $\overrightarrow{\mathcal{J}_{1}}$. 
In (screen on, Wi-Fi registered) mode, the recognition rate achieved by $SenseFlow$ is 60\%. 
However, for the phones with (screen off, Wi-Fi registered), the trajectories are hardly recognised, the recognition rate is only 20\%. The reason is the probe request interval is increased when the Wi-Fi connection of smartphone is set up and the screen is locked (shown in Table~\ref{tb_interval}).

\section{Conclusion}
\label{cond}
In this work, we have investigated \textit{SenseFlow} system to monitor people density and track people flow by using a passive collection of probe requests from their smartphones. We explored the lightweight system architecture and data collection scheme based on the probe request interval. In \textit{SenseFlow}, the probe request overhearing problem is addressed to improve people tracking performance. We measured the effect of smartphone's operational modes and human walking behaviour, and evaluated the tracking accuracy of \textit{SenseFlow} in four typical application scenarios.  

\section*{Acknowledgement}
This research is supported by the MND Research Fund Sustainable Urban Living Grant. 

\bibliographystyle{elsarticle-num}
\bibliography{BibTracking}

\end{document}